%% file: spectroastrometry.tex
\documentclass[iop,twocolumn]{emulateapj}

\usepackage{graphicx,amssymb,amsmath,amsfonts}
\usepackage{color,subfigure}
\usepackage{natbib}
\usepackage[section]{placeins}
\usepackage{cases}
\DeclareRobustCommand{\ion}[2]{%
\relax\ifmmode
\ifx\testbx\f@series
{\mathbf{#1\,\mathsc{#2}}}\else
{\mathrm{#1\,\mathsc{#2}}}\fi
\else\textup{#1\,{\mdseries\textsc{#2}}}%
\fi}

\input{commands.tex}

\newcommand{\rblr}{r_{\rm BLR}}
\newcommand{\tblr}{\theta(\rblr)}
\newcommand{\rkaspi}{r_{\rm RM}}
\newcommand{\tkaspi}{\theta(\rkaspi)}
\newcommand{\rpsf}{r_{\rm PSF}}

\newcommand{\Fobs}{F_{\nu;\,1450{\rm \AA}}}
\newcommand{\Lmono}{L_{1450\AA}}

\newcommand{\Phis}{\Phi^\ast}
\newcommand{\Ss}{S^\ast}
\newcommand{\Phicont}{\Phi_{v}^{\rm cont}}

\newcommand{\strehl}{{\it Strehl}}

\newcommand{\psf}{PSF}

\newcommand{\finst}{f_{\rm inst}}

\newcommand{\vK}{v_{\rm K}}
\renewcommand{\vr}{v_{\rm rot}}
\newcommand{\vrsini}{\vr\sin i}

\newcommand{\blue}{_{\rm blue}}
\newcommand{\red}{_{\rm red}}
\newcommand{\cont}{_{\rm cont}}
\newcommand{\bluecont}{_{\rm blue\,cont}}
\newcommand{\redcont}{_{\rm red\,cont}}

\begin{document}

\title{Spatially Resolving the Kinematics of the $\lesssim100\muas$ Quasar Broad Line Region\\Using Spectroastrometry }
\author{Jonathan~Stern\altaffilmark{1}, Joseph~F.~Hennawi\altaffilmark{1}, J{\" o}rg-Uwe~Pott\altaffilmark{1}}
 \altaffiltext{1}{Max Planck Institut f\"{u}r Astronomie, K\"{o}nigstuhl 17, D-69117, Heidelberg, Germany.}
 
\email{E-mail: stern@mpia.de}

\begin{abstract}
 The broad line region (BLR) of luminous active galactic nuclei (AGN) is a prominent observational signature of the accretion flow around supermassive black holes, which can be used to measure their masses ($\mbh$) over cosmic history. Due to the $\lesssim100\muas$ angular size of the BLR, current direct constraints on BLR kinematics are limited to those provided by reverberation mapping studies, which are most efficiently carried out on low-luminosity $L$ and low-redshift $z$ AGN. We analyze the possibility to measure the BLR size and study its kinematic structure using \emph{spectroastrometry}, whereby one measures the spatial position  centroid of emission line photons as a function of velocity.  We calculate the expected spectroastrometric signal of a rotation-dominated BLR for various assumptions about the ratio of random to rotational motions, and the radial distribution of the BLR gas.  We show that for hyper-luminous quasars at $z < 2.5$, the size of the low-ionization BLR can already be constrained 
with existing telescopes and adaptive optics systems, thus providing a novel method to spatially resolve the kinematics of the accretion flow at $10^3 - 10^4$ gravitational radii, and measure $\mbh$ at the high-$L$ end of the AGN family. With a 30m-class telescope, BLR spectroastrometry should be routinely detectable for much fainter quasars out to $z\sim 6$, and for various emission lines. This will enable kinematic $\mbh$ measurements as a function of luminosity and redshift, providing a compelling science case for next generation telescopes. 
\end{abstract} 


\section{Introduction}\label{sec: intro}

Quasars are the most luminous compact objects in the Universe, with luminosities $L$ reaching up to $10^{48}\ergs$.
Such immense energy outputs are believed to be the result of
gas accretion onto black holes (BHs) with masses of up to $\mbh\approx10^{10}\msun$ (\citealt{Lynden-Bell69}), shining at luminosities reaching their Eddington luminosity ($1.3\times10^{38}\,\mbh/\msun\ergs$). 

Arguably the most distinct property of quasar spectra is the prominence of broad emission lines, with typical widths of $\sim3,000\kms$ (e.g.\ \citealt{VandenBerk+01}), and maximum widths of $10-20,000\kms$ (\citealt{Laor03, Fine+08, SternLaor12a}). 
The ionization structure of the broad line emitting gas, known as the broad line region (BLR), is well described by photo-ionization by the central continuum source (\citealt{DavidsonNetzer79} and citations thereafter).
However, the spatial structure and dynamics of the BLR are not well understood. The large observed velocities suggest that the kinematics of the BLR are dominated by the BH gravity, and that the BLR resides at $\sim10^3-10^4\,\rg$, where $\rg$ is the gravitational radius. 

Assuming the observed BLR velocities are of the order of the local Keplerian velocity,
then a measure of the BLR distance from the BH, $\rblr$, provides a measure of $\mbh$. 
Since luminous quasars are observable to high redshift $z$, then estimates of $\mbh$ in quasars can be utilized to track the buildup of massive BHs over cosmic time. 
In low-$L$ and low-$z$ Active Galactic Nuclei (AGN), the family of objects of which quasars are the luminous subset, $\rblr$ can be measured with reverberation mapping (RM). In RM, one derives $\rblr$ from the time lag between changes in the continuum luminosity and the corresponding variation in the line luminosity (\citealt{BlandfordMcKee82,Peterson93,Peterson+04}).
RM has been widely applied in the past two decades, and has yielded three main results. 
First, the response of \Hb\ to changes in continuum luminosity in single objects suggests that the BLR resides in a narrow range of $r$, 
where the bulk of the \Hb\ emission originates from a dynamical range of $5-10$ in $r$ (\citealt{Maoz+91, Pancoast+13}). 
Second, the characteristic response-weighted $\rblr$ of $\Hb$ scales as $\sim L^{1/2}$ (\citealt{Kaspi+05,Bentz+09b,Bentz+13}), with a typical value of $0.05\pc$ at $L_{45}=1$, where $L_{45}$ is the monochromatic luminosity at 1450\AA\ $\Lmono$ in units of $10^{45}\ergs$.
And third, different broad emission lines are emitted from somewhat different $r$, where generally higher ionization lines originate from smaller $r$ than the Balmer lines. 

The observed value of $\rblr$, its scaling with $L$, and its small dynamical range, are all aptly explained by a combination of two effects which suppress the line emission at $r\ll 0.05\,L_{45}^{1/2}\pc$ and $r> 0.1\,L_{45}^{1/2}\pc$. 
At $r\gtrsim\rsub\approx 0.1\,L_{45}^{1/2}\pc$, where $\rsub$ is the dust sublimation radius, dust grains can survive and hence suppress the line emission (\citealt{NetzerLaor93}).
At $r\ll 0.05\,L_{45}^{1/2}\pc$, the equilibrium of BLR gas pressure with radiation pressure suggested by \cite{Baskin+14a} implies that the gas volume densities are so high ($\gg10^{11}\cm^{-3}$), that emission of the permitted lines is collisionally suppressed. The equilibrium with radiation pressure can also explain the stratification of lines with ionization (see fig.\ 5 and \S4.6 in \citealt{Baskin+14a}).

Do the aforementioned properties of $\rblr$ apply also to high-$L$ and high-$z$ quasars?
RM of a couple of $L\approx10^{47}\ergs$ quasars at $z\approx2$ suggests that the answer is yes (\citealt{Kaspi+07,Chelouche+12}). 
However, applying RM to luminous quasars is problematic since in
high-$L$ quasars $\rblr$ is expected to be on the scale of
light-years, with correspondingly 
long response times. Also, since quasar evolution dictates that high-$L$ quasars are at high-$z$, cosmological time dilation makes the observed time lags even longer. Furthermore, high luminosity quasars are less variable (\citealt{AngioneSmith72,VandenBerk+04}) and therefore have a weaker RM signal. 
The combination of a weak signal with RM timescales of several years in the observed frame makes RM of high-$L$ high-$z$ quasars extremely challenging, and hence to date RM studies of only a couple of such objects have been published. 
Therefore, an independent method to constrain $\rblr$ (and $\mbh$) at high-$L$ and high-$z$ would be a big advantage.

Another open question concerns the kinematic structure  of the BLR, which given that it resides at $\sim10^3-10^4\,\rg$, is likely an integral part of the accretion flow. 
Is the BLR in some kind of ordered flow, such as an extension of the accretion disk, 
or are the motions disordered, 
such as a population of clouds in random virial motion about the BH?
There are some indirect observations which favor an ordered flow, such as 
the rotation of the polarization angle across the \Ha\ spectral profile seen in some objects (\citealt{Smith+05}),
and the distinct variability characteristics of the red and blue line wings which is sometimes observed (\citealt{VeilleuxZheng91}).
Both the rotation of the polarization angle and the distinct variability patterns as a function of velocity suggest that the red-wing emitting gas and the blue-wing emitting gas are spatially distinct, as expected in an ordered flow. 
Also, an ordered flow is favored due to the lack of `spikes' in the broad emission line profile, which are expected if the BLR is composed of individual clouds in a random velocity field (\citealt{Arav+98,Dietrich+99,Laor+06}). 
However, a direct observation which confirms the existence of an ordered velocity field in the BLR remains elusive.

In this study, we analyze the possibility to constrain the BLR size and velocity field in high-$L$ and high-$z$ quasars using spectroastrometry (e.g.\ \citealt{Bailey+98}). 
In spectroastrometry, one measures the position centroid of photons as a function of photon wavelength, which in principle can be pinpointed to an accuracy which is $\sim N_{\rm ph}^{1/2}$ higher than the resolution limit of the telescope, where $N_{\rm ph}$ is the number of photons. 
Extrapolation of the $\rblr-L$ relation mentioned above to the most luminous quasars implies BLR angular sizes of $\approx100\muas$, 
which is a factor of $\sim 10^3$ below the resolution of $\approx50\mas$ of diffraction-limited observations on 8m telescopes using adaptive optics at near-infrared wavelengths.
Therefore, with $\sim 10^{6}$ photons,
the $100\muas$-scale photocenter offset between the red-wing and blue-wing photons can in principle be detected. 
As spectroastrometry is most effective in luminous quasars where $\rblr$ is large and the photon flux is high, it is complementary to RM which is most easily
applied to low-$L$ AGN.  
Additionally, spectroastrometry measures an $r$-weighted function of the BLR (see below), compared to the response-weighted function of the BLR measured by RM, hence the two methods give independent constraints on the size, kinematics, and distribution of material in the BLR. 

\cite{Bailey+98} applied spectroastrometry to seeing limited observations of binary stars and of the narrow line region (NLR) in AGN.
Spectroastrometry was also applied to nuclear gas in the Circinus galaxy (\citealt{Gnerucci+13}), to molecular disks in young stellar objects (\citealt{Pontoppidan+08, Pontoppidan+11, Joergens+13}), and to planetary nebulae (\citealt{Blanco+14}).  
Using Adaptive Optics, \citeauthor{Pontoppidan+11} succeeded in overcoming atmospheric and technical issues and achieved photon-limited angular resolutions of $100-500\muas$, suggesting that resolving the BLR with spectroastrometry is feasible with existing telescopes.

Conceptually similar to BLR spectroastrometry, \cite{Shen12} discusses the possibility to exploit the astrometric position of a variable broad line emission, via imaging with a narrow band filter with wavelength tuned to the broad emission line. As discussed in \cite{Trippe+10}, measuring the 2D-astrometric position is dominated by optical aberrations in the imaging instrument, and current imagers show astrometric biases at the $400\muas$ level, a factor of $>$4 too large for BLR spectroastrometry. Hence, \cite{Shen12} concluded that detecting the astrometric signal of the BLR will be delayed to next-generation telescopes. However, in spectroastrometry the optical abberation bias is circumvented, since all photons practically travel along the same path through the optics (see below), and therefore the astrometric signal can in principle be detected already by existing telescopes. 
Additionally, since the technique suggested by \cite{Shen12} requires a variable BLR, it has the same observational resource requirements as RM, and is therefore practically limited to focus on the same highly and shortly variable low-$L$ low-$z$ objects. 
Therefore, in this paper we limit the discussion to the simpler possibility of applying spectroastrometry to the BLR without any temporal information.

This paper is structured as follows.
In \S2 we calculate the BLR angular sizes for the most luminous quasars at each $z$. 
In \S3 we estimate the expected spectroastrometric signal for a rotation-dominated BLR, for different assumptions on the radial distribution
of the BLR gas and on the ratio of random to rotational motion. 
We present several observational considerations in \S4, 
and simulate the expected spectroastrometric signal with $8{\rm m}$ and $30{\rm m}$-class telescopes in \S5. 
We discuss our results in \S6, and summarize in \S7. 
Throughout the paper, we assume a FRW cosmology with $\Omega$ = 0.3, $\Lambda$ = 0.7 and $H_0 = 70\ \kms$ Mpc$^{-1}$.

\section{The Angular Size of the Broad Line Region}\label{sec: angular size}
\cite{Kaspi+05} derived the relation between $\Lmono$ and $\rkaspi$, the distance of the BLR measured by reverberation mapping of \Hb, at a luminosity range $10^{41}<\Lmono<10^{46}\ergs$. 
To estimate $\rblr$ in spectroastrometry candidates, we extrapolate their
relation to the higher luminosities considered here $\Lmono\gtrsim10^{47}\ergs$:
\begin{equation}\label{eq: Kaspi05}
 \rkaspi = 0.5 \left(\frac{\Lmono}{10^{47}\ergs}\right)^{0.5} \pc  ~~~.
\end{equation}

We can convert eqn.~(\ref{eq: Kaspi05})
to a cosmology-independent relation based on angular size $\theta$ and observed flux. 
Define $\Fobs$ as the flux density at rest-frame wavelength $\lrest=1450\AA$, and defining $\dL$ as the luminosity distance, we have 
\begin{equation}\label{eq: nln1450}
 \Lmono = \frac{\nu_{1450\AA}}{1+z}  \Fobs \cdot 4\pi \dL^2 ~~~.
\end{equation}
Eq.~(\ref{eq: nln1450}), combined with the relation between angular distance and luminosity distance $\rkaspi/\tkaspi =\dL/(1+z)^2$, gives
\begin{equation}\label{eq: tblr vs f_nu}
 \tkaspi = 51 \left(\frac{\Fobs}{10\mJy}\right)^{1/2} \left(1+z\right)^{3/2} \muas ~~~.
\end{equation}
It is convenient to express eqn.~(\ref{eq: tblr vs f_nu}) also with the commonly used $\mi(z=2)$, the apparent $i$-band magnitude after K-correcting to $z=2$ (i.e.\ the apparent magnitude at $\lrest=2500\AA$). Therefore, 
\begin{equation}
\Fobs=0.74\cdot 10^{-0.4(\mi(z=2) + 48.6)} ~~~,
\end{equation}
where the factor of $0.74$ follows from a \cite{Richards+06} quasar template. Together with eqn.~(\ref{eq: tblr vs f_nu}), we get
\begin{equation}\label{eq: tblr vs mi}
 \tkaspi = 22\cdot 10^{\left(15-\mi(z=2)\right)/5} \left(1+z\right)^{3/2} \muas ~~~.
\end{equation}

\begin{figure}
    \includegraphics{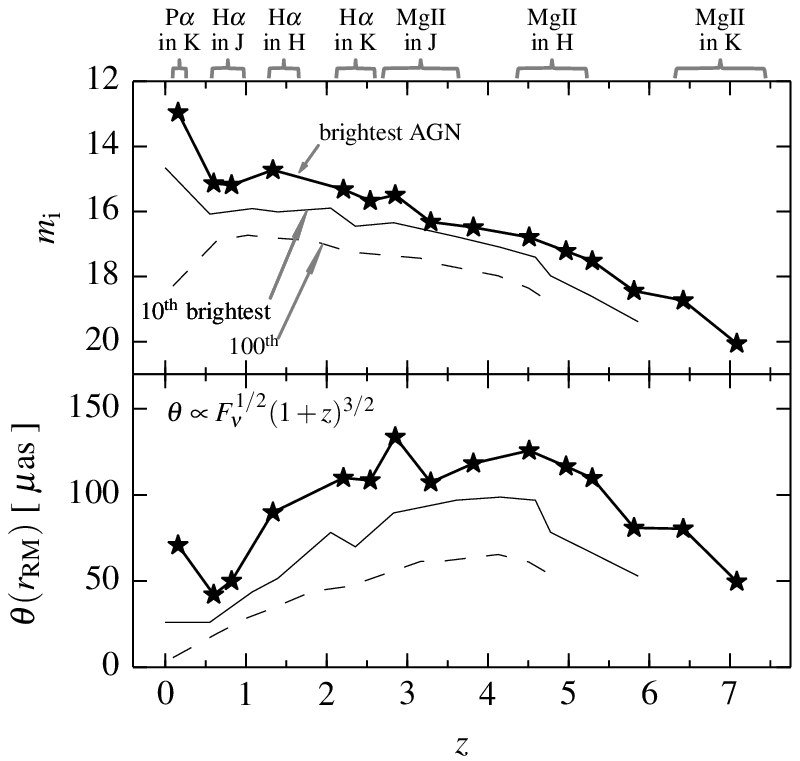} 
    \caption{
    The apparent magnitudes and expected BLR angular sizes of the brightest AGN at each $z$, complied from BQS, SDSS, the $z>5.7$ quasar sample of \cite{Banados+14}, and a sample of local Seyferts. 
    {\bf (Top)} For each 0.5-wide bin in $z$, the values of $\mi$ (K-corrected to $z=2$) of the brightest quasar, the 10\th\ brightest, and the 100\th\ brightest quasar, are shown. 
    The ranges in $z$ where \Pa, \Ha, and \mgiip\ are observed in NIR bands are noted on top. 
    {\bf (Bottom)} The expected BLR angular size of the quasars shown in the top panel, derived from $\Lmono$ by assuming the RM-based relation between $\rkaspi$ and $\Lmono$.
    The implied cosmology-independent relation between $\theta$ and $F_\nu$ (eqn.~\ref{eq: tblr vs f_nu}) is noted. 
    The dependence of the maximum $\tkaspi$ on $z$ is remarkably weak, spanning merely a factor of three over the entire $z$ range. 
}
    \label{fig: angular size}
\end{figure}

We search for the brightest quasars at each $z$ in the Bright Quasar Survey (BQS, \citealt{SchmidtGreen83}), the quasar catalogs of the Sloan Digital Sky Survey 7\th\ and 10\th\ data releases (SDSS-DR7, \citealt{Schneider+10}; SDSS-DR10, \citealt{Paris+14}), and the $z>5.7$ quasar sample compiled by \cite{Banados+14}. 
We supplement this quasar list with local bright Seyferts from \cite{Bentz+09a}, who corrected the AGN continuum emission for the host galaxy contribution, which may be significant in the relatively low luminosity Seyferts. 
We derive $\mi(z=2)$ for SDSS quasars by K-correcting the PSF magnitude of the SDSS-band which is closest to 2500\AA\ in rest-frame wavelength. 
For BQS quasars we derive $\mi(z=2)$ from the B-magnitude, and for the high-$z$ quasars we use the J-band magnitudes, if available, or otherwise the y-band magnitude. 
The luminosity of the local Seyferts is based on the luminosity at 5100\AA\ listed in table 9 of \cite{Bentz+09a}. 
All K-corrections are performed assuming a \cite{Richards+06} quasar template. 
We divide the quasars into bins of $0.5$ in $z$, and sort them by brightness. The brightest\footnote{None of these quasars appears in the list of lensed quasars compiled by \cite{VeronCettyVeron10}.}, 10\th\ brightest, and 100\th\ brightest quasar in each $z$-bin are shown in Figure \ref{fig: angular size}. 
For reference, we note the ranges in $z$ where \Pa, \Ha, and \mgiip\ are observed in NIR bands. 
As expected, $\mi(z=2)$ of the brightest quasars generally decreases with $z$. 
However, due to the factor of $(1+z)^{3/2}$ in eqn.~(\ref{eq: tblr vs mi}), the dependence of the maximum $\tkaspi$ on $z$ is remarkably weak, spanning merely a factor of three ($50-150\muas)$ over the entire $z$ range. 

\section{The spectroastrometric signal of the BLR}\label{sec: spectroastrometric signal}

\cite{Chen+89} and \citeauthor{ChenHalpern89}\ (1989, hereafter CH89)
derived the spectral profile of a broad emission line originating from
a rotating disk, including both special and general relativistic
effects to first order in $\rg/r$.
The two main parameters of their
model are the distribution of line emission as a function of $r$, and
the amount of line broadening at each location of the disk, due to unordered motions which could be sourced by turbulence or other physical processes.
Below, we discuss the physical and observational constraints on these two parameters.
Here, we use the CH89 formulation to derive the expected spectroastrometric signal of a rotating BLR as a function of these two parameters. 

We note that the flat disk assumed by CH89 is probably an oversimplification even for a rotation-dominated BLR, since the BLR characteristic height above the disk mid-plane may change with $r$. Therefore, our analysis is based on the assumption that as long as the dominant motion is rotation, the CH89 formulation should give a reasonable approximation of the expected spectroastrometric signal.

\begin{figure} 
    \includegraphics{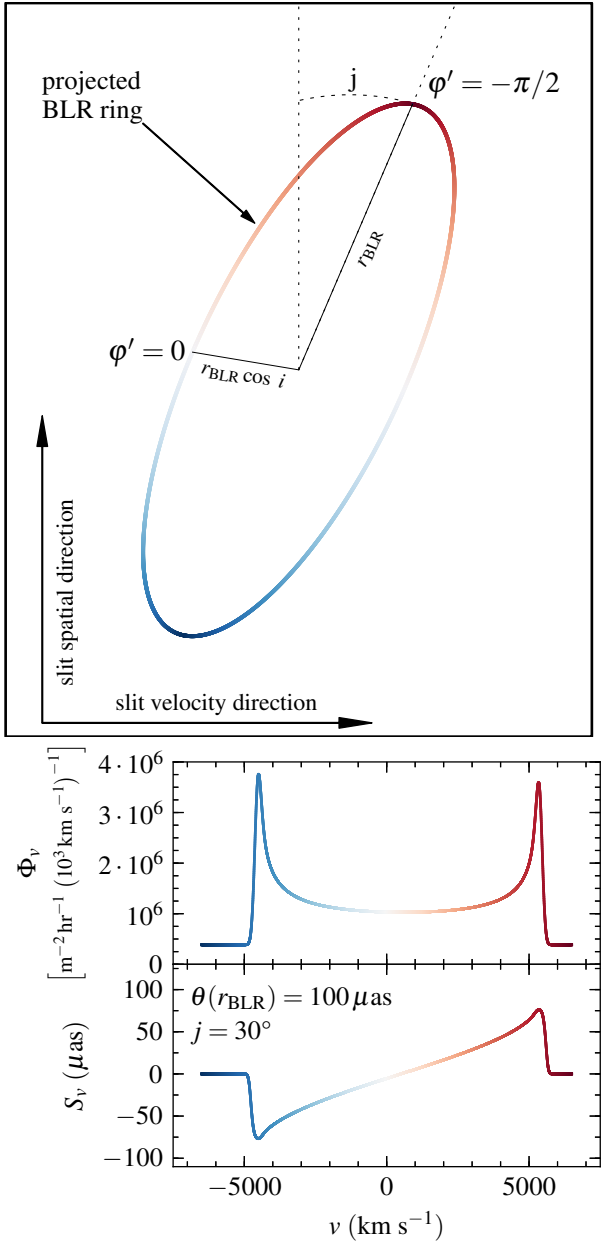} 
    \caption{{\bf (Top)} An illustration of a spectroastrometric observation of a simplified BLR, which originates from a single $r=\rblr$ and has
      negligible local line broadening. 
      The ellipse depicts the BLR ring, with coordinate $\varphi'$, projected onto the plane of the sky. Color denotes line-of-sight velocity. 
      The physical size of the ellipse axes, up to relativistic effects, are $\rblr$ and $\rblr\cos i$. 
      The angle between the spatial direction of the slit and the major axis of the projected ring is defined as $j$. 
      {\bf (Middle)} The implied emission-line profile of the rotating ring on top of the underlying quasar continuum, for an assumed $\vr=10\,000\kms$, $\sin i=0.5$, and ${\rm EW}=26\,000\kms$.  Normalization assumes an \Ha\ line of a $L=10^{48}\ergs$ quasar at $z=2$. 
      {\bf (Bottom)} The implied photocenter offsets of the BLR, assuming no contamination by non-BLR sources, for the $\tblr$ and $j$ noted in the panel. 
      The centroid of the reddest line photons are offset from the centroid of the bluest line photons by $150\muas$. 
}
    \label{fig: setup}
\end{figure}

Following CH89, we define $r$ and $\varphi'$ as the polar coordinates in the disk frame. 
We assume that the orientation on the sky of the spatial direction 
of the telescope slit is at an angle $j$ from the major axis of the projected BLR ellipse.
These definitions are pictured in the top panel of Figure \ref{fig: setup}, which depicts an idealized BLR, which originates from a single $r=\rblr$.

We define $\Phis_v(r, \varphi')\d r\d\varphi' ~~\left[\cm^{-2}\s^{-1} \left(\kms\right)^{-1}\right]$ as the observed flux density of photons per unit velocity which originate from the disk coordinate $(r,\varphi')$.
Similarly, the photocenter position of these photons along the spatial direction of the slit is defined as $\Ss_v(r,\varphi')\ \left[\muas\right]$. 
The value of $\Ss_v(r,\varphi')$ can be calculated from the impact parameter of a photon at infinity $b$, which equals (eqn.~6 in CH89) 
\begin{equation}\label{eq: b}
b=r\left(1-\sin^2 i\cos^2\varphi'\right)^{1/2}\left(1+ O(\frac{\rg}{r})\right) ~~~,
\end{equation}
where $i$ is the inclination of the disk normal to the line of sight. The first term in eqn.~(\ref{eq: b}) is the Newtonian projection of $(r,\varphi')$ on the sky, and the second term is a correction due to light bending by the central mass. 
Projecting $b$ on the slit spatial direction yields, with some trigonometry, 
\begin{eqnarray}\label{eq: Sphs long}
 \Ss_v(r,\varphi') = r\left(\sin j \cos i \cos \varphi' + \cos j \sin \varphi'\right)\left(1+ O(\frac{\rg}{r})\right) ~. \nonumber \\
\end{eqnarray}

The observable photocenter position at velocity $v$, $S_v$, is the photon-flux-average of $\Ss_v(r,\varphi')$: 
\begin{equation}\label{eq: Sphs}
 S_v = \frac{\iint \Ss_v(r,\varphi') \Phis_v(r, \varphi')\d \varphi'\d r}{\Phi_v} ~~~,
\end{equation}
where $\Phi_v$ is the observed photon flux density:
\begin{equation}\label{eq: Nphs}
\Phi_v = \iint \Phis_v(r, \varphi')\d \varphi'\d r ~~~.
\end{equation}
Now, since $\Phis_v(r, \varphi') = \Phis_v(r, \pi-\varphi')\left(1+ O(\rg/r)\right)$ (see CH89), then by summing the contribution from $\varphi'$ and $\pi-\varphi'$ to the integral in eqn.~(\ref{eq: Sphs}), the term in eqn.~(\ref{eq: Sphs long}) with $\cos \varphi'$ cancels out to zeroth order in $\rg/r$, and we get:
\begin{eqnarray}
 \frac{1}{2}\left[ \Ss_v(r,\varphi') \Phis_v(r, \varphi') + \Ss_v(r,\pi-\varphi') \Phis_v(r, \pi-\varphi')\right]  =  \nonumber \\
    r\cos j \sin \varphi'\Phis_v(r, \varphi')\left(1+ O(\frac{\rg}{r})\right) ~~~.
\end{eqnarray}
Hence, eqn.~(\ref{eq: Sphs}) simplifies to  
\begin{equation}\label{eq: S_v0}
 S_v = \cos j\cdot \frac{\int r\d r\int \d \varphi' \sin \varphi' \Phis_v(r, \varphi')\left(1+ O(\frac{\rg}{r})\right)}{\Phi_v}
\end{equation}

The light-bending corrections to $S_v$ are of the order of $O(\rg/\rblr)\lesssim 10^{-3}$, and are likely weaker than 
the inaccuracy of the model due to the assumption that the BLR is a flat rotating disk, so it is not clear whether these corrections are observable\footnote{Note that relativistic corrections of orders $O((\rg/\rblr)^{1/2})$ and $O(\rg/\rblr)$ to the line {\it profile} have been observed (CH89; \citealt{Tremaine+14}).}. 

The observed $S_v$ will be diluted by the quasar continuum, which at optical and ultraviolet frequencies originates from the accretion disk. The continuum emission will have a negligible spectroastrometric signal, and therefore will dilute $S_v$ of the broad line by a factor of $\Phi_v / (\Phi_v+\Phicont)$, where $\Phicont$ is the photon flux density of the continuum\footnote{Equivalent to eqn.~4 in \cite{Pontoppidan+08}.}. Therefore, the final form of eqn.~(\ref{eq: S_v0}) is
\begin{equation}\label{eq: S_v}
 S_v = \cos j\cdot \frac{\int r\d r\int \d \varphi' \sin \varphi' \Phis_v(r, \varphi')\left(1+ O(\frac{\rg}{r})\right)}{\Phi_v + \Phicont}
\end{equation}

\subsection{A Simplified BLR}
In order to explore the dependence of $S_v$ on the disk properties, we begin by calculating $S_v$ for a simplified BLR which originates from a single thin ring
$r=\rblr$
and has negligible local line-broadening, i.e.\ $\sigma \ll \vr$, where $\sigma$ is the characteristic velocity of the local line-broadening mechanism, and $\vr$ is the rotational velocity. Under these conditions, 
$\vr=(G\mbh/\rblr)^{1/2}$.  
We relax both assumptions below. 
In the figures, we use the fully relativistic expression for $\Phis_v(r,\varphi')$ from CH89\footnote{The value of $\Phis_v(r,\varphi')$ is equal to the integrand in eqn.~7 in CH89, divided by the observed photon energy.},
and eqn.~(\ref{eq: S_v}) to calculate $S_v$. 
However, since the relativistic corrections to $\Phis_v(r,\varphi')$ have a weak effect on the derived $S_v$, for increased readability we repeat here the Newtonian expressions for $\Phis_v$.
In this simplified BLR, $\Phi_v^{\ast}$ is equal to
\begin{equation}\label{eq: Phi1}
 \Phi_v^{\ast}(r, \varphi')\propto\delta(r-\rblr)\delta(\sin\varphi'-\frac{v}{\vr\sin i})
\end{equation}
The implied $\Phi_v$ is shown in
the second panel of Fig.\ \ref{fig: setup}, assuming $\vr=10\,000\kms$, $i=30\degree$, and a total broad line photon flux of  $4\times10^6~{\rm photons} \hr^{-1} \m^{-2}$.
This broad line flux is appropriate for the broad \Ha\ line of a $L=10^{48}\ergs$ quasar at $z=2$, assuming an equivalent width (EW) of $570\AA$ compared to the AGN continuum (\citealt{SternLaor12a}). We note that \citeauthor{SternLaor12a} found that EW(\Ha) is independent of $L$ at $L<10^{46}\ergs$, and therefore we assume that the same EW pertains at $L\gg10^{46}\ergs$. The implied EW in velocity units is $26,000\kms$, and the continuum flux density is derived from this assumed EW. 
The units are chosen for easy conversion into the number of photons expected in an observation. 

Using eqn.~(\ref{eq: Phi1}) in eqn.~(\ref{eq: S_v}), the implied $S_v$ for the simplified BLR is 
\begin{equation}\label{eq: Sv1}
 \left. S_v\right|_{v\leq\vrsini} = \frac{\Phi_v}{\Phi_v+\Phicont}\cdot\rblr \cos j \frac{v}{\vrsini} ~~~,
\end{equation}
which is shown in the bottom panel of Fig.\ \ref{fig: setup}, assuming $j=30\degree$ and $\tblr=100\muas$. 
The $v=5000\kms$ photons are offset from the $v =-5000\kms$ by $\approx2\rblr\cos j\cdot\Phi_v/(\Phi_v+\Phicont)=150\muas$.

\subsection{Local Line Broadening}\label{sec: turbulence}

\begin{figure}
    \includegraphics{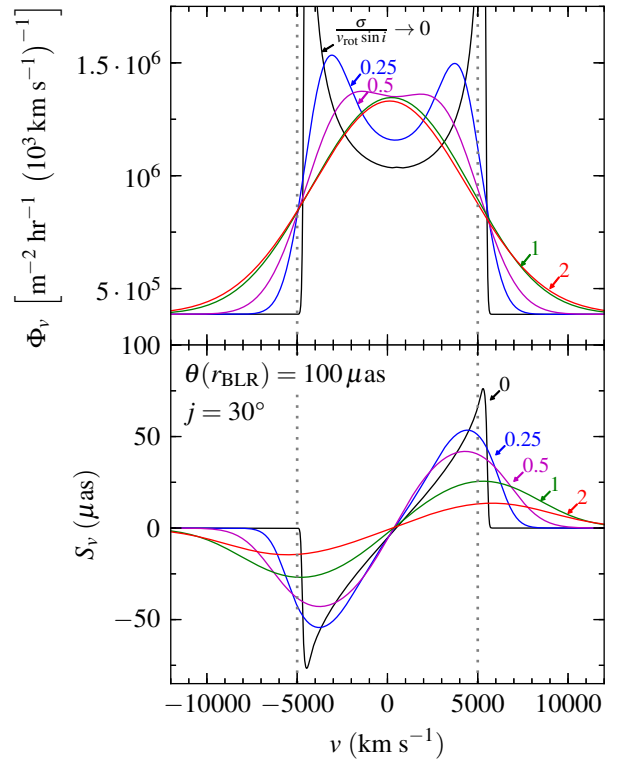} 
    \caption{
    The emission line profile (upper panel) and photocenter offsets (lower panel) of a BLR ring with local line broadening. 
    At each location in the ring, the emitted photon wavelength distribution is assumed to be a Gaussian with width $\sigma$. 
    The black lines assume negligible broadening as in Fig.~\ref{fig: setup}, while other plotted lines assume the noted value of $\sigma / (\vrsini)$. 
    As $\sigma / (\vrsini)$ increases, the two peaks in the line profile merge into a single peak, and the maximum value of $|S_v|$ decreases. 
}
    \label{fig: turbulence}
\end{figure}

We now relax the assumption that the local line broadening at each location in the disk is negligible. Therefore,
\begin{equation}\label{eq: Phi2}
 \Phi_v^{\ast}(r, \varphi')\propto
\delta(r-\rblr) {\rm e}^{-\frac{\left(\vrsini\right)^2}{2\sigma^2}\left(\sin \varphi' - \frac{v}{\vrsini}\right)^2} ~~~.
\end{equation}
In Figure \ref{fig: turbulence} we plot $\Phi_v$ and $S_v$ for different $\sigma/(\vrsini)$. 
In order to match the emission line-profile of a putative observation, 
$\vr$ is chosen so that the line width remains constant in the different models.
For convenience, we characterize the line width by the Half Width at Half Maximum ($\hwhm$). 

Figure \ref{fig: turbulence} illustrates that as $\sigma/(\vrsini)$
increases, $\Phi_v$ loses its double-peaked shape and the
maximum spectroastrometric signal $|S_v|$ decreases.  This behavior occurs
because as $\sigma/(\vrsini)$ increases, photons from a wider range of $\varphi'$ 
contribute to a given value of $v$, thus smoothing the spectral profile and $S_v$. 

What is the value of $\sigma$ expected in the BLR?
There are some indirect observations which suggest that $\sigma \sim \vrsini \sim 0.5\,\vr$\footnote{The typical value of $\sin i$ is expected to be $\approx0.5$ in the standard AGN unification model, where the broad line region is visibile only if the quasar is viewed at pole-on inclinations.}.
In most quasars the BLR profiles do not exhibit a double-peaked profile (\citealt{Strateva+03}),
which is expected when $\sigma < 0.5\,\vrsini$ (see top panel of Fig.\ \ref{fig: turbulence}). 
This lack of a double-peak is sometimes attributed to the BLR being emitted from a wide range of $r$, however this explanation is disfavored by the RM results which suggest that the BLR originates from a narrow range in $r$ (see \S\ref{sec: intro}). Therefore, the single-peaked profile typically observed is more likely due to $\sigma/\vrsini$ of order unity. 
Furthermore, \cite{Osterbrock78} argued that the statistics of BLR line widths suggest $\sigma/\vr\sim0.4$, 
which is also roughly consistent with the inferred BLR covering factor (CF) of $\sim0.3$ (\citealt{Korista+97}), assuming that $\sigma/\vr$ is a measure of the height-to-diameter ratio of the BLR disk. 
Due to the above arguments, for the purpose of estimating the strength of the spectroastrometric signal we will henceforth assume $\sigma/(\vrsini) \approx 1$. 
The bottom panel of Fig.\ \ref{fig: turbulence} shows that $\sigma/(\vrsini) \approx 1$  implies $S_{\pm\hwhm}\approx\pm0.25\,\tblr$.

\subsection{Distribution in $r$}\label{sec: r-distribution}

\begin{figure}
    \includegraphics{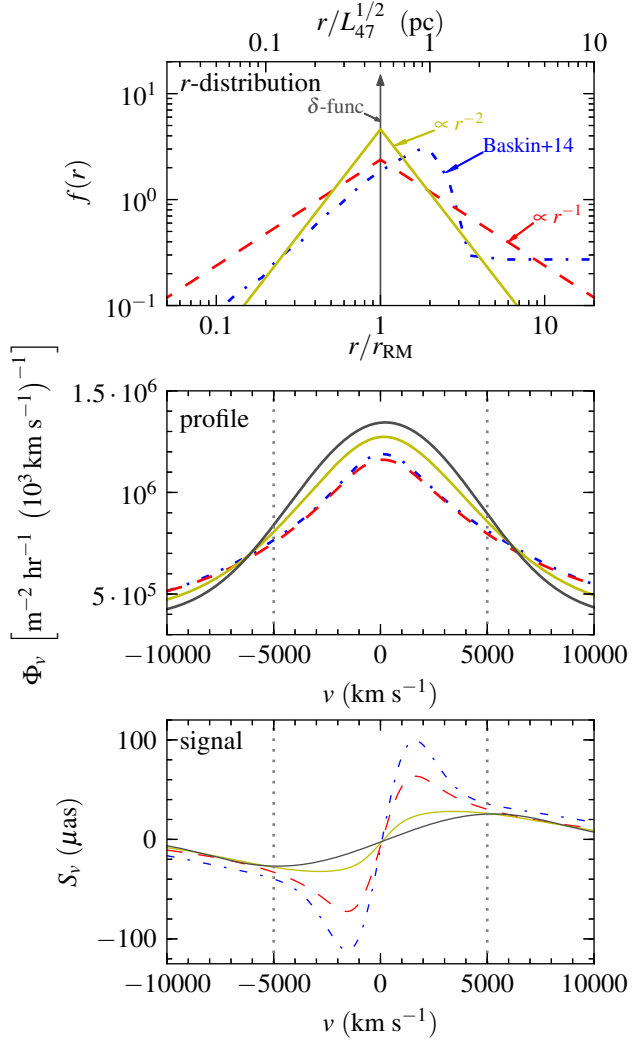}
    \caption{
    {\bf (Top)} Possible radial distributions of the line emission. 
    The black arrow is a $\delta$-function (thin ring), as used in Figs.~\ref{fig: setup} and \ref{fig: turbulence}. 
    The red solid line and the green dashed line are parameterizations of $f(r)$ which are consistent with RM studies, where $f(r)$ peaks at $r=\rkaspi$ and drops at higher and lower $r$. The top axis notes the physical scale for a quasar with $L_{1450\AA}=10^{47}\ergs$. 
    The blue dash-dotted line is a physically-motivated \Ha-emission distribution from Baskin et al. (2014, fig.\ 5 there), where the line emission drops at $r>\rsub=2\rkaspi$ due to absorption of the incident ionizing photons by dust grains, and also drops at $r\ll\rkaspi$ due to collisional suppression of \Ha.
    {\bf (Middle)} The line spectral profiles for the different $f(r)$ shown in the top panel, assuming $\sigma/(\vrsini)=1$ and $\hwhm=5000\kms$. 
    The spectral profiles are similar to each other. 
    {\bf (Bottom)} The expected photocenter offsets for the different $f(r)$. 
    At $|v|<\hwhm$, $S_v$ is highly sensitive to the relatively weak emission from $r\gg\rkaspi$. 
    At $|v|\approx\hwhm$, $S_v$ only weakly depends on the assumed $f(r)$. 
}
    \label{fig: r-distribution}
\end{figure}

We now relax the idealized assumption that the BLR originates from a single $r$, and consider a general spatial distribution of the broad line emitting gas, parameterized by $f(r)$, the line emission per unit $\log r$:
\begin{equation}\label{eq: Phi3}
 \Phi_v^{\ast}(r, \varphi') = \frac{f(r)}{r}{\rm e}^{-\frac{\left(\vrsini\right)^2}{2\sigma^2}\left(\sin \varphi' - \frac{v}{\vrsini}\right)^2} 
\end{equation}
What is the expected shape of $f(r)$?
As noted in \S\ref{sec: intro}, RM studies suggest that the BLR emission originates from a small dynamical range in $r$, i.e. $f(r)$ is strongly peaked around $r=\rkaspi$. We therefore, parameterize $f(r)$ as a double power law $f(r)\propto r^{\pm\alpha}$, where the index is positive at $r<\rkaspi$ and negative at $r>\rkaspi$, and calculate $S_v$ for $\alpha = 1$ and $2$. These $f(r)$ are shown in the top panel of Figure \ref{fig: r-distribution}, together with a $\delta$-function (thin ring) for comparison with Figs.\ \ref{fig: setup} and \ref{fig: turbulence}.
We assume that $f(r)$ covers the range $0.03\,\rkaspi - 30\,\rkaspi$. 
Additionally, we also consider a physically motivated $f(r)$, where $f(r)$ drops at $r>2\rkaspi=\rsub$ due to absorption of the ionizing photons by dust grains,
 and also drops at $r<\rkaspi$ due to collisional suppression of the line emission. 
This $f(r)$ was calculated for different lines by \cite{Baskin+14a}, assuming a constant BLR CF per unit $\log r$. 
Since below we estimate $S_v$ for \Ha, we use the \citeauthor{Baskin+14a}\ calculation of $f(r)$ of \Ha, which is similar in shape to $f(r)$ of \Hb\ seen in their fig.\ 5\footnote{\citeauthor{Baskin+14a}\ plot $f(r)$ in terms of the line EW, assuming a CF per unit $\log\,r$ of $0.3$. We use their BLR model with solar metallicity and an ionizing spectral slope of $-1.6$.}. We refer the reader to \citeauthor{Baskin+14a}\ for the details of the calculation. 

The middle and lower panels of Fig.\ \ref{fig: r-distribution} show the implied $\Phi_v$ and $S_v$, for the different $f(r)$ shown in the top panel.
As above, we assume $\tkaspi=100\muas$, $i=j=30\degree$, an EW of a typical \Ha\ ($=26\,000\kms$), and $\vr$ at $r=\rkaspi$ such that $\hwhm=5000\kms$. We assume $\vr(r)\propto r^{-1/2}$, and based on the arguments
in the previous section, $\sigma(r)/\vr(r)\sin\,i = 1$. 
The middle panel shows that $\Phi_v$ of the different $f(r)$ are similar, up to somewhat more extended line wings in the $f(r)\propto r^{\pm1}$ and \citeauthor{Baskin+14a}\ distributions compared to the $f(r)\propto r^{\pm2}$ and $\delta$-function distributions. This similarity is expected since all assumed $f(r)$ drop sharply when $r$ is significantly larger than or significantly smaller than $\rkaspi$. 
In contrast, the shape of $S_v$ seen in the lower panel differ significantly for the different $f(r)$. At $v=2000\kms$, for example, $S_v=25\muas$ for $f(r)\propto r^{\pm2}$, compared to $S_v=60\muas$ for $f(r)\propto r^{\pm1}$ and $S_v=90\muas$ for the $f(r)$ calculated by \citeauthor{Baskin+14a}
However, at $|v|\approx\hwhm=5000\kms$, the dependence of $S_v$ on $f(r)$ is significantly weaker than at $v=2000\kms$, where $S_{v=\hwhm}$ is in the range $25-37\muas$ for all considered $f(r)$.

Why is $S_v$ at $v<\hwhm$ so sensitive to emission from large $r$?
As seen in eqn.~(\ref{eq: S_v}), $S_v$ is a photon-weighted average of the contribution to $S_v$ from each $r$. 
Eqn.\ (\ref{eq: Sv1}) demonstrates that the contribution to $S_v$ from a specific $r$ scales as $\propto r\vr^{-1}\propto r^{3/2}$, for $v\leq\vr(r)\sin i$. 
While eqn.~(\ref{eq: Sv1}) is derived assuming $\sigma\ll\vrsini$, the bottom panel of Fig.\ \ref{fig: turbulence} shows that it is also roughly a correct description of $S_v$ for general $\sigma$, up to multiplication by a constant which depends on $\sigma$.  
Therefore, in the case where $f(r)\propto r^{-1}$ at $r>\rkaspi$, this $r^{3/2}$ term implies that $S_v$ is dominated by photons from the largest $r$ with significant emission at $v$, despite that these large-$r$ photons have a negligible contribution to the line profile. This effect is even more pronounced for the physical $f(r)$ calculated by \cite{Baskin+14a}, in which more photons originate from $r>10~\rkaspi$ than in the $f(r)\propto r^{\pm1}$ case. 

To conclude, the spectroastrometry signal at $v<\hwhm$ is very
sensitive to $f(r)$ at $r>\rkaspi$, and therefore can be used to probe
the line emission from this regime, as further discussed in
\S\ref{sec: discussion}.  At $v$ comparable to the \hwhm, $S_v$ probes
$\rkaspi$, i.e.\ $S_v$ probes the scale at which the bulk of the BLR emission is produced. 

\subsection{Contaminants} 

In a real observation, some of the photons with wavelengths spanned by the observed broad line originate from non-BLR sources, and therefore we need to account for their affect on $S_v$. 
We have already referred to the effect of quasar continuum photons, which dilute $S_v$, and can be accounted for by dividing the observed $S_v$ by the fraction of line photons at each $v$ (see \S\ref{sec: spectroastrometric signal}).
In this section, we discuss additional non-BLR contaminants and how to account for their effect on $S_v$. 

\subsubsection{Narrow emission lines}\label{sec: NLR}

Photons from the NLR originate from $r\gg\rblr$, and hence any asymmetries on NLR scales (as seen in fig.~5 of \citealt{Bailey+98}) may dominate $S_v$ at $v$ with narrow line emission.
The largest relevant $r$ is $\rpsf$, the physical size spanned by the \psf, since emission from $r>\rpsf$ can likely be filtered out during the data reduction phase of the spectroastrometric observation. 
For an assumed ${\rm PSF}=70\mas$ we get $\rpsf \approx 10^3\,\rblr$ ($\rpsf=600\pc$ at $z=2$). 
Now, since in the calculation of $S_v$ (eqn.~\ref{eq: S_v}) there is a factor of $r$ in the integrand, 
NLR emission which originates from $10^3\,\rblr$ may dominate $S_v$, 
even if the NLR flux density is only $\sim0.1\%$ of the broad line flux density. 
This is the same argument made in \S\ref{sec: r-distribution} for why weak BLR emission from large scales can dominate $S_v$. Hence, photons from wavelengths near narrow emission lines will need to be disregarded when measuring the photocenter of the BLR.

For example, at a broad \Ha\ luminosity of $10^{44}\ergs$, which is equivalent to $L=10^{46}\ergs$ (\citealt{SternLaor12a}), the mean fluxes of the narrow \Ha, \oi, \sii\ and \nii\ lines are $0.3-3\%$ of the broad \Ha\ flux (fig.~3 in \citealt{SternLaor12b}, fig.~1 in \citealt{SternLaor13}). The flux {\it density} ratio of \nii\ and the narrow \Ha\ is $\sim10$ times this value for a typical NLR-to-BLR \fwhm\ ratio of $0.1$, while \oi\ and \sii, which reside on the broad \Ha\ wings, have even higher flux density ratios. 
Therefore, since the NLR to BLR flux density ratios is $\gg0.1\%$ at $v$ with these strong narrow lines, the relevant $S_v$ will likely be dominated by the NLR.

We search also for weaker lines in the \cloudy\ (\citealt{Ferland+13}) model of the NLR described in \cite{Stern+14}\footnote{We use the \cite{Stern+14} NLR model with an ionizing slope of $-1.6$, solar metallicity, and $r=600\,L_{47}^{1/2}\pc$, which corresponds to $r=1000\,\rblr\approx\rpsf$, where the NLR contribution to $S_v$ peaks.}. 
We find three additional lines which may be strong enough to dominate $S_v$. The lines are \siii, \arv, and \hei, at $v_\Ha=-11\,500,~5840,$ and $5270\kms$, and have fluxes which are $0.08$, $0.03$, and $0.01$ times the flux of the narrow \Ha, respectively. 
The latter two lines can be seen as small bumps in the high resolution spectra of NGC~4151 (fig.~1 in \citealt{Arav+98}). 
Photons near these lines may also have to be disregarded. 
In \S\ref{sec: simulation} below we demonstrate that the emission from even weaker lines will not significantly affect $S_v$. 

We note that the above estimates for the NLR strength is likely an upper limit, since spectroastrometry candidates have $L\gg10^{46}\ergs$, while the narrow to broad flux ratio decreases with $L$ as $\sim L^{-0.3}$, probably due to the decrease in NLR covering factor with increasing $L$ (\citealt{SternLaor12b}). 
Additionally, some of the narrow line emission may originate from $r>\rpsf$ and therefore can be filtered out.

\subsubsection{Continuum emission from the host galaxy, distant scattering media, and dust grains}\label{sec: host}

At optical wavelengths, the polarized flux density in unobscured AGN is typically $0.5-5\%$ of the total flux density (\citealt{Smith+02}), suggesting that some of the observed emission is scattered emission. The scattering medium may reside on scales $\gg\rblr$, and therefore, based on the same reasoning used above for the NLR, asymmetries in this scattering medium can in principle dominate $S_v$. 
However, because a distant scatter views the BLR as a point source, the opening angle covered by the scatterer will not be a function of location in the BLR, and hence not a function of velocity. Since the scattering efficiency is a weak function of $\lambda$, and a $v=10\,000\kms$ emission line spans merely $\d\lambda/\lambda=0.03$, the scattering efficiency is also a weak function of velocity. Therefore, we expect the contribution of the scattered flux to $S_v$ to be roughly constant across the emission line and the flanking continuum.
We can account for the effect of a distant scatterer by interpolating $S_v$ from the continuum flanking the broad line profile, and subtracting out the result. This approach will also account for the effect of the host galaxy emission on $S_v$, since any asymmetry in the host galaxy is also likely to be a weak function $\lambda$.

We note that the rotation of the polarization angle across the \Ha\ profile found by \cite{Smith+02,Smith+05} suggests the existence of scattering media also at $r\gtrsim \rblr$. Such nearby scatterers will not dominate $S_v$.

If one measures the spectroastrometric signal of a rest-frame IR emission 
line such as \Pa, then the dust IR continuum can affect $S_v$. The
$S_v$ profile of the dust continuum is also expected to be a weak
function of $\lambda$, and therefore can be accounted for using the
same interpolation method noted above for accounting for scattering media. 
However, since the dust continuum is a significant fraction of
the flux at IR wavelengths, in contrast with the host and scattered
flux, any second-order dependence of $S_v$ on $\lambda$ which is not
accounted for may significantly affect $S_v$, and therefore may pose a
problem for performing spectroastrometric on rest-frame IR lines.

\section{Observational Considerations}

\subsection{Systematics}\label{sec: noise}

We note that most of the atmospheric and instrumental systematic errors relevant to imaging astrometry are irrelevant to BLR spectroastrometry, since in BLR spectroastrometry one measures the astrometric position of the broad emission line relative to the adjacent continuum, rather than in any absolute sense. Specifically, the red and blue sides of the broad emission line are expected to be offset from the continuum in different directions (Figs.~\ref{fig: setup} -- \ref{fig: r-distribution}). 
Most systematic effects on the photon centroid are expected to be a weak function of $\lambda$, since the line and continuum photons travel in the exact same way through the atmosphere and instrument, and are only
split in the wavelength domain. Therefore, to mimic a spectroastrometric signal, any instrumental systematic needs to vary so strongly with $\lambda$ that within $\d\lambda/\lambda \approx 0.03$ it produces a relative offset between the red and blue wings of the emission line. 
Any systematic which is a weak function of $\lambda$ can be accounted for by interpolating the
centroid of the continuum flanking the emission line, a method which
accounts also for the effect of astrophysical continuum sources such as the host galaxy or scattering media
(\S\ref{sec: host}).
The error of this interpolation is limited
by the photon count of the continuum, and is estimated below. 

On the other hand, instrumental systematics which are a strong function of $\lambda$ may be a limiting factor, rather than
the number of photons collected. 
\cite{Pontoppidan+11} showed that rotating the slit by $180\deg$ can mitigate such systematics which are related to flat-fielding and removal of telluric lines. 
This method is based on the fact that the spectroastrometric offset is a real offset between line photons on the sky, and therefore a $180\deg$ rotation of the slit flips the direction of the spectroastrometry offset in detector coordinates, while most instrumental systematics would be fixed in
detector coordinates. 
By applying this $180\deg$ flip, \citeauthor{Pontoppidan+11}\ reached {\it photon-limited}
accuracies of $100-500\muas$, i.e.\ at an accuracy of $100\muas$ their observations were still free of systematic effects. 
We therefore find it reasonable that systematics are, or can be reduced to $\sim25\muas$, the accuracy level required for BLR spectroastrometry.

An additional method to differentiate between a true signal and an instrumental systematic would be to test the predicted $j$-dependence of the spectroastrometric signal (eqn.~\ref{eq: S_v}), using multiple slit orientations. In order for an instrumental systematic to mimic the spectroastrometric signal it would to
have to behave exactly the same way with rotation of the slit.

\subsection{Choice of Target and Emission Line}

The choice of an optimal target and emission line for
spectroastrometry is motivated by four main factors. The angular size
of the emitting region, the photon flux of the line, the strength of
the non-BLR contaminants listed in the previous section, and the
observed wavelength of the line. 
The $\Ha$ line scores highly given
all of these considerations, since it is the strongest optical
emission line, it is emitted from the outer part of the BLR
(\citealt{Bentz+10}), and its wavelength is near $1\mic$ where the
accretion disk and dust continuum emission are at a
minimum. Furthermore, at $z=2-2.5$ where $\tkaspi$ is near its peak
value (Fig.~\ref{fig: angular size}), H$\alpha$ is redshifted into
the near-IR K-band where adaptive-optics works best. 

Also, we note that the idea of an ordered rotational velocity field in the BLR is best established for low-ionization lines, such as the Balmer lines. Higher ionization lines appear also to have an outflowing wind component (\citealt{Richards+11}, see further discussion in \S\ref{sec: discussion}).
Hence, interpreting the spectroastrometic signal of high-ionization lines will likely be more complicated than interpreting the signal of low ionization lines. 

Given the above arguments, below we simulate the expected spectroastrometry signal for \Ha. We consider also the prospect of
measuring $S_v$ on \mgiip, which falls in the NIR bands at $z=3-7.5$, and also $S_v$ of \Pa\ in the local quasar 3C~273, the
brightest quasar on the sky.

\subsection{Choosing the Slit Orientation}\label{sec: j} 
Since $S_v\propto\cos j$ (eqn.~\ref{eq: S_v}), the required number of photons to achieve a given $\snr$ scales as $\cos^{-2}j$, where $j$ is not known a-priori.
In the context of measuring the spectroastrometric signal of stellar disks,
\cite{Pontoppidan+08}
suggested an observing strategy where one observes the target at three slit
orientations rotated by $60\degree$ from one another.
Since $\left(\cos^2 (j) + \cos^2 (j-60) + \cos^2 (j-120)\right)/3 = 0.5$ for all $j$, than the required observing time with this technique is twice the observing time required for the $j=0$ case.
Therefore, below we assume $j=0$, and multiply the required observing time by a factor of two to account for the unknown $j$. 

We note that even if one performs a spectroastrometric observation with an Integral Field Unit (IFU), which maps the sky into 2D pixels and preforms simulatenous spectroscopy in each pixel, rotation of the unit is still likely to be required. An IFU maps one spatial direction continuously onto the detector, while the other spatial direction is mapped discontinuously. 
Constructing the spectroastrometic signal along the spatial dimension which is mapped continuously onto the detector is tantamount to constructing the signal in a long-slit observation. 
However, constructing the spectroastrometic signal along the dimension which is mapped discontinously requires modelling the projection of this spatial dimension onto the detector to the level of $10^{-3}$ of a pixel, as required for BLR spectroastrometry, and is therefore unlikely to be feasible. Therefore, using an IFU has no clear advantage over using a long-slit.

\section{Simulated Spectroastrometry Signal}\label{sec: simulation}

In this section we simulate the spectroastrometric signal for a $z=2$
target on 8m and 39m telescopes, and then consider measurements of
targets at other redshifts.

\begin{figure} 
    \includegraphics{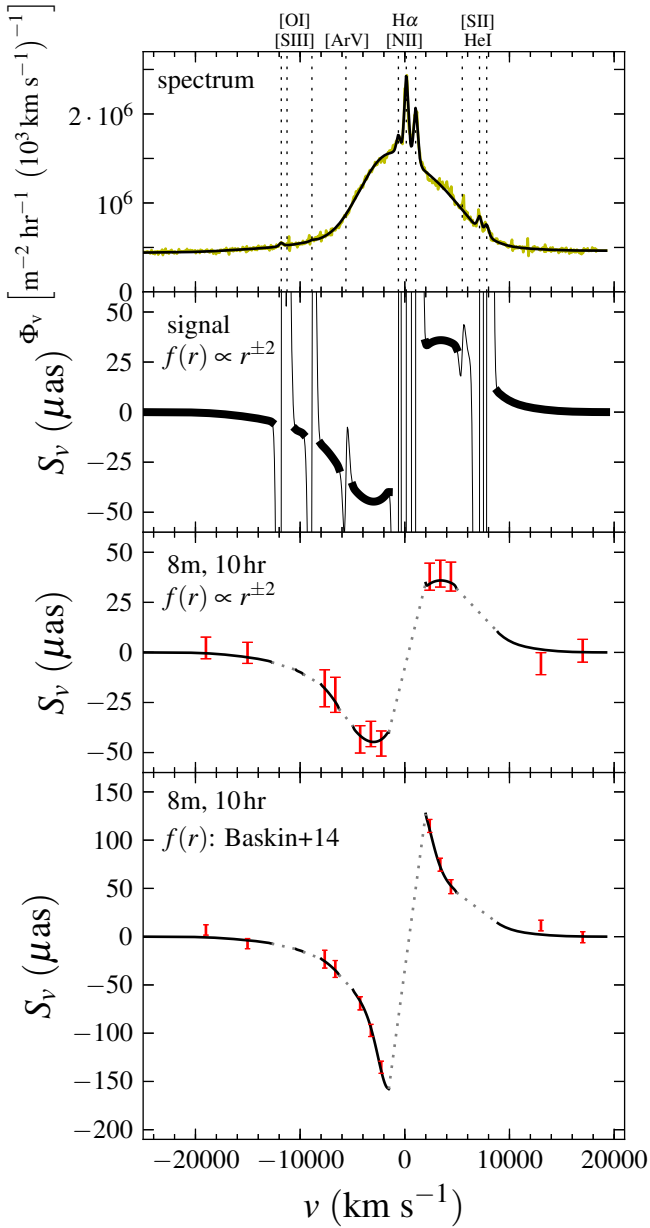}
    \caption{The simulated spectroastrometric signal of the broad \Ha\ in J1521+5202, a $\Lmono=10^{47.6}\ergs$ quasar at $z=2.2$. 
	     {\bf (Top panel)} The yellow line plots the normalized spectrum of J1540-0205, used as a template for J1521+5202. The black line plots the fit to the continuum, broad \Ha, and the noted narrow lines (marked by vertical dotted lines). 
	     {\bf (2$^{\rm nd}$ panel)} The expected $S_v$, assuming $\sigma/(\vrsini)=1$ and the $f(r)\propto r^{\pm2}$ shown in Fig.~\ref{fig: r-distribution}. The narrow lines likely have asymmetry on large scales, therefore at $v$ with strong NLR emission $S_v$ is on the scale of the PSF, which is beyond the plotted axis. The weak \arv\ and \hei\ lines have a comparable contribution to $S_v$ as the broad \Ha. At $v$ which are not dominated by narrow lines (thick line), the broad \Ha\ and quasar continuum dominate $S_v$. The photocenter at $v=4000\kms$ is offset from the photocenter at $v=-4000\kms$ by $85\muas$. 
	     {\bf (3$^{\rm rd}$ panel)} The solid line marks the expected $S_v$ at $v$ where the NLR does not dominate $S_v$. 
	     The error bars are the simulated centroid statistical uncertainties, for spectral bin sizes of $1000$ and $4000\kms$ in the emission line and continuum, respectively. We assume a $10\hr$ integration on an AO-assisted 8m telescope, with $\strehl=0.4$ and $\finst=0.2$. 
	     {\bf (Bottom panel)} The expected $S_v$ and centroid uncertainties assuming the physically-motivated $f(r)$ shown in Fig.~\ref{fig: r-distribution}.
}
    \label{fig: expected signal}
\end{figure}

\subsection{The H$\alpha$ Line of SDSS~J1521+5202}
SDSS~J152156.48+520238.5 (hereafter J1521+5202) is the most luminous SDSS quasar where
\Ha\ falls in the K-band (Fig.~\ref{fig: angular size}). J1521+5202 has $\Lmono=10^{47.6}\ergs$,
$\tkaspi=111\muas$, and $z=2.208$. 
The $\fwhm$ of its \mgiip\ line is
$9300\kms$ (\citealt{Shen+11}).  
As the NIR spectrum of J1521+5202 is not available, we use as a template the spectrum of SDSS~J154019.57-020505.4 (hereafter J1540-0205), a
$\Lmono=10^{45.4}\ergs$ quasar at $z=0.3$ with $\fwhm(\Ha)=9730\kms$. The two spectra should be similar since the $\fwhm$ of the Balmer lines is correlated with the $\fwhm$ of \mgiip\ (\citealt{Shen+08}). 
We multiply the observed photon flux of J1540-0205 by the photon flux ratio of the two quasars,
$10^{47.6}/10^{45.4}\cdot(\dL(z=2.2)/\dL(z=0.3))^2\cdot(1+2.2)/(1+0.3)=3.1$.
The normalized spectrum is shown in the top panel of Figure
\ref{fig: expected signal}, together with the broad \Ha, narrow lines
and continuum fits described in \cite{SternLaor12a}, which we utilize
below. We note that since J1521+5202 is more luminous than J1540-0205,
the host galaxy is expected to be relatively weaker, as are the narrow
emission lines (\citealt{SternLaor12b}).  
We determine the flux of the
weak \siii, \arv, and \hei\ lines from the flux of the narrow \Ha\ and the model flux
ratios mentioned in \S\ref{sec: NLR}. 
These lines are too weak to be observable in the SDSS spectrum.

In order to estimate $S_v$, the broad emission line is modeled as a
rotating disk with $\sigma/(\vrsini)=1$ and either $f(r)\propto
r^{\pm2}$, or the  $f(r)$ based on \cite{Baskin+14a}.
We assume for simplicity that the angular size of the
\Ha-emitting region is $\tkaspi$, though the true $\theta$ may be
higher since $\rkaspi$ was determined on \Hb, and \Ha\ is sometimes
observed to have longer RM lags than \Hb\ (\citealt{Bentz+10}).  We set
$\vr(\rkaspi)\sin i$ of the model such that the $\fwhm$ in the model profile
equals the observed $\fwhm$ of $9730\kms$, i.e.\ $2\vr(\rkaspi)\sin i =
7400\kms$ for the $f(r)\propto r^{\pm2}$ case and $2\vr(\rkaspi)\sin i
= 11\,000\kms$ for the $f(r)$ from \citeauthor{Baskin+14a} 
We normalize the model $\Phi~(\equiv \int\Phi_v\d v)$ to give the same value of the observed $\Phi$. 
We note that at $|v|<5000\kms$, the observed $\Phi_v$ profile of the
broad \Ha\ differs by up to 20\% from the $\Phi_v$ calculated by the
BLR model. 
However, since the goal of this study is to estimate $S_v$ rather than provide a model which accurately fits the observed emission line, we disregard this difference. 

In order to estimate $S_v$ of the narrow lines, we assume the worst-case scenario where the narrow lines originate from $r=\rpsf=600\pc$ (assuming an 8m-telescope diffraction limited \psf\ of $70\mas$), where the contribution to $S_v$ is maximal (\S\ref{sec: NLR}).
For simplicity, we assume the NLR emission also originates from a turbulent rotating disk with $\vrsini$ chosen to fit the observed narrow lines \fwhm\ of $450\kms$.
The total photon flux of the NLR model is chosen to fit the observed photon fluxes of the strong emission lines (\Ha, \oi\, \nii, and \sii), and the model photon fluxes of the weak emission lines (\siii, \arv, \heii). 
Finally, quasar continuum photons are assumed to originate from $r=0$. 

The second panel in Fig.\ \ref{fig: r-distribution} shows $S_v$ versus
$v$ for $f(r)\propto r^{\pm2}$. Thin lines mark $v$ values which are
contaminated by emission from narrow lines.  The NLR spectroastrometric
signal is on the scale of the PSF, which is beyond the plotted axis. Note that the
weak signal of \hei\ justifies our choice in \S\ref{sec: NLR} to
disregard narrow lines with a flux which is lower than the flux of
\hei. The thick line marks the range of $v$ where the broad \Ha\ and
quasar continuum dominate $S_v$, and only these $v$ are used further
in the analysis.  The photocenters at the blue and red wings are
offset by up to $85\muas$ from each other.

To estimate the expected error on the measured photocenter, we assume a $t=10\hr$ observation on an $8{\rm m}$ telescope. 
Using the specifications of the Nasmyth Adaptive Optics System (NAOS) and COud{\' e} Near Infrared CAmera (CONICA) as guidance\footnote{See user manual at http://www.eso.org .}, 
we assume a collecting area of $A=38{\rm m}^2$, a diffraction-limited $\fwhm_{\rm \psf}=70\mas$, 
a \strehl\ ratio of $0.4$, and an instrument photon collecting efficiency of $\finst=0.2$. 
For the observed $\Phi_v\simeq10^6~[\m^{-2}\hr^{-1}\left(10^3 \kms\right)^{-1}]$
at half maximum of the broad \Ha\ (top panel of Fig.~\ref{fig: r-distribution}), the implied number of collected photons per $1000\kms$ 
is 
\begin{equation}\label{eq: Nph}
 \frac{\d N_{\rm ph}}{\d v} = \Phi_v \cdot A \cdot t \cdot \strehl \cdot \finst \cdot 0.5 
            = 15\times10^6 ~\left(10^3 \kms\right)^{-1} ,
 \end{equation}
where the factor of $0.5$ reduction arises from the slit orientation as discussed in \S\ref{sec: j}.

The value of $\d N_{\rm ph}/\d v$ computed in eqn.~(\ref{eq: Nph}) implies a statistical $1\sigma$ photocenter positioning error $\epsilon_v$ of 
\begin{equation}\label{eq: epsilon}
 \epsilon_v = \frac{\fwhm_{\rm \psf}}{2.35\, \left(\frac{\d N_{\rm ph}}{\d v}\right)^{1/2}} = 7.7 \muas ~~~.
\end{equation}
The third panel in Fig.\ \ref{fig: expected signal} shows the expected errors in the photocenter measurements, for bins of size $\Delta v=1000\kms$ in the red and blue wings, and bins of size $\Delta v=4000\kms$ for the red and blue continuum regions. 
To simulate an observation, the center of the simulated errorbars is offset from the expected signal (solid line) by a random offset chosen from a Gaussian distribution with dispersion $\epsilon_v$. 
Fig.\ \ref{fig: expected signal} demonstrates that if systematics can be reduced to the level of a few times the value of $\epsilon_v$ calculated in eq. \ref{eq: epsilon}, then the photocenter offset of the red wing from the blue wing is detectable with existing telescopes. 
For comparison, \cite{Pontoppidan+11} reached photon-limited uncertainties of $100\muas$ in the context of stellar disks, but this precision level was set by the number of photons collected rather than by systematic errors.

\begin{figure} 
    \includegraphics{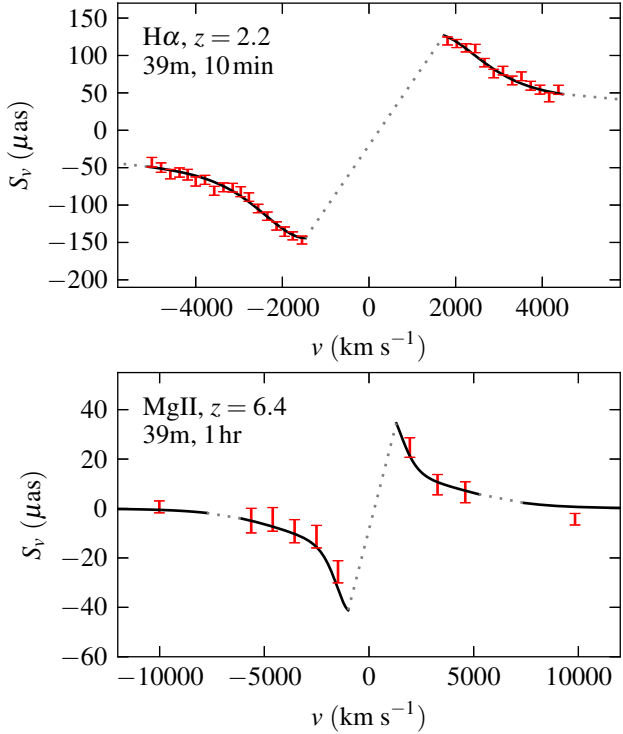}
    \caption{Simulated spectroastrometric signal with a 39m telescope. As in the bottom panels of Fig.~\ref{fig: expected signal}, the solid line denotes the expected $S_v$ at velocities in which the BLR and quasar continuum dominate $S_v$. Error bars denote the expected statistical uncertainty in the photocentroid measurement. In both panels, $f(r)$ from \cite{Baskin+14a} is assumed. 
    {\bf (Top panel)} The signal expected on the broad \Ha\ of J1521+5202, the object which appears also in Fig.~\ref{fig: expected signal}. 
    The higher photon flux and smaller PSF in 39m telescopes compared to 8m telescopes enables constraining $S_v$ to an order of magnitude higher velocity resolution, with a fraction of the observing time. 
    {\bf (Bottom panel)} The signal expected on the broad \mgiip\ line of SDSS J114816.64+525150.3, a $z=6.4$ quasar. 
 }
    \label{fig: 39m}
\end{figure}

In the bottom panel of Fig.\ \ref{fig: expected signal}, we show the expected spectroastrometric signal assuming the $f(r)$ based on \cite{Baskin+14a}. 
As expected from Fig.\ \ref{fig: r-distribution}, at $v<\hwhm$ the maximum expected $S_v$ is a factor of four larger than when assuming $f(r)\propto r^{\pm2}$, and the relative errorbar sizes are correspondingly smaller.

In the top panel of Figure \ref{fig: 39m}, we show the expected spectroastrometric signal of J1521+5202 for a $10\min$ observation on a next-generation $39\m$ telescope, zoomed-in on the emission line for clarity. We assume that $A\propto d^2$, $\fwhm_{\rm \psf}\propto d^{-1}$, a \strehl\ of $0.4$ (e.g.\ \citealt{Clenet+10}), and $\finst=0.4$ (e.g.\ \citealt{Davies+10}). 
Hence, eqs.\ \ref{eq: Nph} and \ref{eq: epsilon} imply that $\epsilon\propto d^{-2}$, and that the required observing time to reach a certain value of $\epsilon$ scales as $d^{-4}$. We therefore decrease the spectral bin sizes to $200\kms$. With a next generation telescope, one can derive $S_v$ to a high velocity resolution, and therefore yield strong constraints on the BLR kinematics.

\subsection{Spectroastrometry Estimates for Additional Quasars}

In this section, we estimate the statistical $\snr$ of the
spectroastrometric signal of the most luminous quasars at different
$z$.  To this end, we define $\snr(S\red-S\blue)$, to be the
photocenter offset between the red and blue wings of a broad emission
line, which is an aggregate $\snr$ that effectively combines the data
points shown in Figs.\ \ref{fig: expected signal} and \ref{fig:
  39m}. For each wing, we include all photons with velocities $0 < |v| < \fwhm$, excluding velocities which might be contaminated by narrow line emission. 
We address quasars in which \Pa, \Ha, or \mgiip\ fall in one
of the NIR transmission windows, and require $\lrest$ is at least $10\,000\kms$ away from
the edge of the atmospheric window, in order to adequately sample the
quasar continuum and avoid significant telluric absorption.
These quasars are listed in Table \ref{tab: snr}, which appears in the Appendix. 
The \snr\ of less luminous quasars can then be estimated by noting
that
\begin{equation}\label{eq: snr vs. L}
 \snr(z) \propto \frac{\tkaspi}{\epsilon_v} \propto \frac{L^{1/2}}{N_{\rm ph}^{-1/2}} \propto L
\end{equation}
where we used eqs.\ \ref{eq: Kaspi05} and \ref{eq: epsilon} for the dependence of $\tkaspi$ and $\epsilon_v$ on $L$, respectively, and emphasize that eqn.~(\ref{eq: snr vs. L}) is only valid at a fixed redshift.
We assume a \strehl\ ratio of $0.2$, $0.4$, and $0.4$, for the J, H and K bands, respectively, and $\finst=0.2~(0.4)$ for all bands for an 8m (39m) telescope\footnote{Instrument throughput based on NACO and MICADO (Multi-AO Imaging Camera for Deep Observations, \citealt{Davies+10}) specifications.}.

Potential narrow line contamination near \mgiip\ and \Pa\ is derived in a similar fashion as for narrow lines near \Ha\ (\S\ref{sec: NLR}).
Narrow lines with a large enough flux to significantly effect $S_v$ near \mgiip\ 
include the narrow \mgiip\ doublet, 
a \heii\ line at $v_\mgiip=-7000\kms$, an \ariv\ line at $v_\mgiip=6000\kms$, and a \mgv\ line at $v_\mgiip=6100\kms$. Similarly, near \Pa\ there are the \Brd\ and \Bre\ lines, three He lines at $v_\Pa=-255,\,-1055$ and $-1855\kms$, and an \feii\ line at $v_\Pa=2140\kms$. All photons with $v$ within $750\kms$ of these lines are excluded from the analysis. We note that some of the mentioned lines are permitted lines, and therefore should also exhibit broad line emission, however these broad lines are likely too weak to affect $S_v$ significantly.

The rotating disk parameters of the model are the same as in the previous section, with $\vr(\rkaspi)$ and the normalization chosen so the model reproduces the estimated values
 of $\fwhm$ and $\Phi$ listed in Table \ref{tab: snr} in the Appendix. Since not all targets have available NIR spectra, we estimate $\fwhm$ and $\Phi$ from observed emission lines, as detailed in the Appendix. 
For \mgiip, we assume that the total flux is contributed from the two lines in the doublet (separated by $770\kms$) according to a $2:1$ ratio. Therefore, at a given $v$ the expected centroid is the photon-flux-weighted centroid of the individual lines in the doublet.
 
The value of $\snr(S\red - S\blue)$ is the number of standard deviations from which $S\red - S\blue$ deviates from a straight line interpolation of the photon centroids of the continuum bins. It is equal to
\begin{equation}\label{eq: snr red - blue}
 \snr(S\red - S\blue) = \frac{S\red - S\blue}{\left(\epsilon\red^2 + \epsilon\blue^2 + \epsilon\cont^2\right)^{1/2}}
\end{equation}
where $\epsilon\blue$ and $\epsilon\red$ are the statistical errors on the photocenters of blue and red wing photons, respectively, calculated from the expressions in eqs. \ref{eq: Nph} and \ref{eq: epsilon}. 
The error on the continuum level, $\epsilon\cont$, is equal to the error on the evaluation of a straight line interpolation from the data points in the blue and red continuum regions:
\begin{equation}\label{eq: eps_cont}
 \epsilon\cont = \frac{v\red - v\blue}{v\redcont - v\bluecont} \left( \epsilon\redcont^2 + \epsilon\bluecont^2\right)^{1/2} ~~~,
\end{equation}
where $\epsilon\bluecont$ and $\epsilon\redcont$ are the statistical errors on the photocenters of each continuum bin ($\fwhm<|v|<2\,\fwhm$), and the continuum flux is calculated from the emission line EW noted in Table \ref{tab: snr}.
The values of $v\red$ and $v\blue$ are the photon-flux average $v$ of the red and blue wing bins, respectively, while $v\redcont=-v\bluecont=1.5\,\fwhm$ are the average velocities of the continuum bins. 
We note that in the $\snr$ calculations shown in Fig.\ \ref{fig: SN vs. z}, $\epsilon\cont^2 / (\epsilon\red^2 + \epsilon\blue^2) = 0.07 - 0.3$, and therefore including $\epsilon\cont$ in the calculation does not significantly affect $\snr(S\red - S\blue)$. 
However, in a real observation the amount of continuum photons may be limited if the emission line falls near the edge of an atmospheric window, which will increase $\epsilon\cont$, and hence decrease $\snr(S\red - S\blue)$.

The expected $\snr(S_{\rm red} - S_{\rm blue})$, for $8{\rm m}$ and $39{\rm m}$ telescopes, are shown in Figure \ref{fig: SN vs. z}. 
At $1<z<2.5$, a night on an $8{\rm m}$ is sufficient to achieve $\snr(S_{\rm red} - S_{\rm blue}) \approx 10$, even when assuming $f(r)\propto r^{-2}$. Assuming the \citeauthor{Baskin+14a}-based $f(r)$ increases the expected $\snr$ by a factor of $\approx2$, while an hour on a $39{\rm m}$ telescope increases the expected $\snr$ by a factor of $\approx10$ compared to $10$ hours on an $8{\rm m}$. Fig.~\ref{fig: SN vs. z} suggests that with next-generation telescopes, the spectroastrometic signal can be detected for quasars as distant as $z=6.5$. As an example, in the bottom panel of Fig.~\ref{fig: 39m} we plot a simulated observation of the \mgiip\ line in SDSS~J114816.64+525150.3, a $z=6.43$ quasar discovered by \cite{Fan+03}. 

\section{Discussion}\label{sec: discussion}
\subsection{The Assumption that the BLR is a Rotating Disk}

For the purpose of considering ordered motion, we assumed that the BLR has a flattened geometry and that its kinematics are dominated by rotation. These assumptions are supported by several observations of the low-ionization BLR gas, which we briefly summarize below. 

\begin{figure} 
    \includegraphics{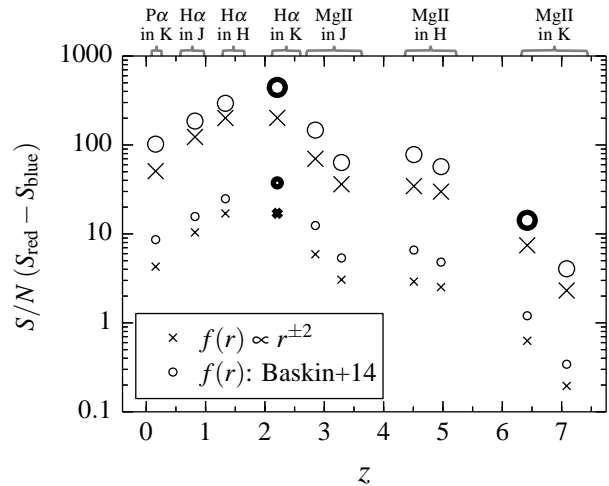}
    \caption{
    Expected S/N for spectroastrometry candidates at different $z$. Candidates are chosen from the most luminous quasars at each $z$ (Fig.~\ref{fig: angular size}), where a strong low-ionization emission line falls in one of the NIR bands. 
    The y-axis is the S/N of the centroid offset of the blue and red wings of the emission line (eqn.~\ref{eq: snr red - blue}), for $10\hr$ on an 8m telescope (small symbols), and $1\hr$ on a 39m telescope (large symbols). 
    The marker type denotes the assumed radial distribution of the line emission (Fig.~\ref{fig: r-distribution}), and bold markers denote the examples shown in Figs.~\ref{fig: expected signal} -- \ref{fig: 39m}.
    The spectroastrometric signal of the BLR in $1<z<2.5$ quasars is detectable at $\snr>10$ with existing telescopes, while next-generation telescopes will be able to observe the spectroastrometric signal up to $z=6.5$. 
}
    \label{fig: SN vs. z}
\end{figure}

Evidence for a flattened geometry comes from the AGN unification picture. According to the the standard AGN paradigm, in unobscured AGN (known as type 1 AGN, including quasars) the accretion disk is viewed
at angles near face-on. When an AGN disk is viewed edge-on, the accretion disk and broad line emission are obscured by a geometrically thick dusty `torus' (\citealt{Antonucci93}). 
\cite{Maiolino+01} and \cite{Gaskell09} argued that if the BLR is spherically symmetric, then a fraction of  the quasars, equal to the BLR covering factor of $\sim30\%$, should be viewed through the BLR clouds. Such BLR gas along the sightline would produce Lyman-edge absorption and low ionization broad absorption lines (LoBALs).
However, only 1\% of quasars are LoBALs (\citealt{Trump+06}), in contrast with the prediction of a spherically symmetric BLR. 
Hence, a more likely picture is that the low-ionization BLR resides predominantly near the disk-plane in a flattened geometry, along line of sights where the quasar would be viewed as obscured.

As for our assumption of a rotating BLR, rotation is preferred over an outflow-dominated BLR, since in a scenario where the BLR is an outflow off the face of the disk, the redshifted and blueshifted photons will originate from opposite locations along the minor-axis of the projected BLR ellipse, in contrast to along the major-axis in the rotating BLR scenario (see top panel of Fig.\ \ref{fig: setup}). This geometry implies a shorter response time of the blue wing compared to the response time of the red wing, which is ruled out in velocity-mapped RM studies of low ionization lines (\citealt{Maoz+91, Grier+13, Pancoast+13}), albeit for low-$z$ low-$L$ AGN.
Further support for the rotating BLR assumption comes from the mean redshift of the broad \Hb\ relative to systemic in SDSS quasars, which is consistent with the relativistic redshift expected in a rotating BLR (\citealt{Tremaine+14}),
and from the analysis of the microlensing signal of the quadruply lensed quasar HE~0435-1223 (\citealt{Braibant+14}). 

However, the high-ionization broad lines likely have a significant outflowing component. Evidence for an outflow can be seen in the line profile of \civp~$\lambda1450$ (\citealt{BaskinLaor05,Richards+11}), and in the high-ionization states typically observed in broad absorption lines (\citealt{Hamann97}), which are predominantly blueshifted, suggesting an outflow. 
Since in an outflow scenario the spatial separation of the redshifted and blueshifted gas is along the minor axis, which is smaller than the major axis by a factor of $\cos i$, the expected $S_v$ should also be correspondingly smaller (assuming the same $\rblr$). Since this signal reduction is relatively small, the spectroastrometric signal should be detectable also in an outflowing BLR.

Note that as the rotational motion in the BLR decreases relative to the random motion, $S_v$ is reduced as shown in the bottom panel of Fig.\ \ref{fig: turbulence}. 
In J1521+5202, for example, the $\snr(S\red-S\blue)=17$ expected for $10\hr$ on an 8m telescope (Fig.\ \ref{fig: SN vs. z}) is reduced to $\snr(S\red-S\blue)<5$ if $\sigma/(\vrsini)>3$.
Therefore, even a null detection of the spectroastrometric signal would place an interesting upper limit on the fraction of the BLR motion that is ordered. 

\subsection{$\mbh$ Estimates}

Above we have demonstrated that by measuring the BLR photocenter offset $S_v$, one can derive $\rblr$. 
Given a measurement of $\rblr$ and an estimate of the Keplerian velocity $\vK$ based 
on the line width, one can derive $\mbh = \rblr \vK^2 / G$, analogous to RM-studies. 
Therefore, since any uncertainty in the $\rblr$ measurement propagates to the $\mbh$ estimate, it is interesting to understand the implied uncertainty in $\rblr$ given a measurement of $S_v$. 

For a given $\rblr$, different values of $\sigma$ imply changes of a factor of two in $S_v$ (Fig.\ \ref{fig: turbulence}). Also, Fig.\ \ref{fig: r-distribution} shows that while different radial distributions of the line emission can change the $S_v$ profile significantly, the range of $S_v$ at $|v|=\hwhm$ for different $f(r)$ spans merely a factor of $1.5$. So, in the worst case scenario where nothing is known about $\sigma$ and $f(r)$, for a given measurement of $S_v$ at $|v|=\hwhm$ one can estimate $\rblr$ to within a factor of $\sim3$.
However, this uncertainty can be reduced by constraining $\sigma$ from the line profile, and by constraining $f(r)$ using the measurements of $S_v$ at $|v|<\hwhm$. 
For the purpose of estimating $\mbh$, these uncertainties on the $\rblr$ estimate should be added to the uncertainty on $\vK$ due to the unknown inclination $i$ of the disk plane to the line of sight. 
The total implied accuracy to which $\mbh$ can be derived is an important topic for future study. 

Furthermore, $\sigma$ and $f(r)$ are likely to be similar or the same for all BLRs. In this case, one can tie the spectroastrometry
measurements of the BLR to lower-luminosity sources which have been reverberation mapped, potentially removing this source of `noise'. 
With next-generation telescopes, one can potentially perform spectroastrometry on objects which are low luminosity enough to be also reverberation mapped.

\subsection{8m-class Telescopes, 30m-class Telescopes and Space Telescopes}

Fig.\ \ref{fig: SN vs. z} shows that existing telescopes can detect the BLR spectroastrometric signal in \Ha\ in the most luminous quasars at $1<z<2.5$. Therefore, with existing telescopes one can use spectroastrometry to estimate the $\mbh$ of the most luminous quasars at the peak of the quasar epoch. Also, the measured $\rblr$ can be compared to the extrapolation of the $\rblr-L$ relation deduced by RM studies on lower luminosity AGN (eqn.~\ref{eq: Kaspi05}).

In \S\ref{sec: simulation} we show that the required integration time to reach a given $\snr$ scales as $d^{-4}$, where $d$ is the telescope diameter. Therefore, with next generation 30m-class telescopes, the required integration times are lower by a factor of $(30/8)^4=200$ than with 8m telescopes.
Also, the smaller \psf\ of larger telescopes implies that reducing the systematics to a given angular precision is likely less challenging.
 
The relatively short integration times required in 30m telescopes can be utilized to observe a large sample of quasars (from eqn.~\ref{eq: snr red - blue} and Fig.\ \ref{fig: SN vs. z}, reaching $\snr=10$ on quasars with $L_{1450}=10^{47}\ergs$ at $z\approx2$ requires $7-15$ minutes on a 39m).
Thus, one can explore the dependence of $S_v$ on additional BLR parameters: e.g.\ the BLR \fwhm, the EW of the lines, $\mbh$ and $\lledd$, whether the high-ionization BLR shows signs of an outflow (\citealt{Richards+11}), and whether the quasar spectrum shows Broad Absorption Lines, which might be some proxy for orientation. 
The high photon flux can also be utilized to achieve a higher resolution in $v$, as shown in the top panel of Fig.\ \ref{fig: 39m}, and thus further constrain the kinematic model of the BLR. 
Additionally, with next generation telescopes one can explore fainter quasars, such as quasars as distant as $z=6.5$ (Fig.\ \ref{fig: SN vs. z}), 
and fainter emission lines.
This opens the possibility to perform spectroastrometry on multiple lines simultaneously, at specific $z$ where multiple lines fall in the NIR atmospheric windows (say $\ciii~\lambda1909$ in J and $\mgiip$ in H at $z\sim5$). Thus, one can constrain the BLR structure as a function of the ionization level of the line-emitting gas. 

Another interesting possibility is to perform BLR spectroastrometry using the James Webb Space Telescope (JWST, \citealt{Gardner+06}), which has a diameter of 6.6m and allows full coverage at $0.6-5\mic$ without the issue of telluric absorption. Since $\strehl=1$ in a space telescope, compared to $\strehl=0.4$ assumed above for an 8m telescope, the observing time requirements are comparable $( (6.5/8)^{-4}\times 0.4=0.9 )$. 
The continuous wavelength coverage can allow observing multiple emission lines for the same quasar. Utilizing the Hubble Space Telescope (HST) for BLR spectroastrometry may also be possible, since the increased observing time due to the relatively small telescope diameter of 2.4m is partially offset by the lower diffraction limit in the UV. We differ a more thorough analysis of the possibility to perform BLR spectroastrometry with space telescopes to future work. 

\subsection{Comparison with the Interferometry Results of \cite{Petrov+12}}
One can also use differential interferometry to measure the photocenter offset of the BLR relative to the continuum. 
\cite{Petrov+12} applied this approach to the P$\alpha$ line of 3C~273, and measured an angular radius of $\gtrsim400\muas$. As noted there, this angular radius is larger by a factor of $\gtrsim3$ than the radius implied by RM of \Ha. 
We suspect that this apparent discrepancy between $\rblr$ measured by interferometry and $\rblr$ measured by RM is due to the high sensitivity of the photocenter to emission from large $r$ (Fig.\ \ref{fig: r-distribution} and \S\ref{sec: r-distribution}). Fig.\ \ref{fig: r-distribution} demonstrates  that when a small fraction of the line photons come from $r\gg\rkaspi$ (i.e. when $f(r)\propto r^{\pm1}$ or when using the physically-motivated $f(r)$ calculated by \citealt{Baskin+14a}), then the average $S_v$ at $0<v<\hwhm$ increases by a factor of $3-4$ compared to when assuming a $\delta$-function radial distribution. This increase is consistent with the discrepancy found by \citeauthor{Petrov+12}

We note that a key advantage of spectroastrometry over interferometry is sky coverage. 
Currently interferometers require co-phasing of the telescopes with a bright star or AGN ($\lesssim 12\mag$ in the NIR), which needs to be at a small angular separation on the order of the isoplanatic angle at the science wavelength (typically $<20\arcsec$, \citealt{Esposito+00}). 
Only a handful of nearby quasars (Fig.\ \ref{fig: angular size}) are bright enough such that a nearby star is unnecessary. 
In contrast, spectroastrometry can benefit from the $\sim50\%$ sky coverage provided by Laser-Guide-Star Adaptive Optics (e.g.\ \citealt{Diolaiti10}).

\subsection{Constraints on Torus Models}

As first noted by \cite{NetzerLaor93}, the line emission from gas at $r>\rsub$ which is exposed to the quasar ionizing radiation is suppressed, due to absorption of the ionizing photons by dust grains, rather than absorption of the ionizing photons by gas particles. 
This suppression, seen as a drop at $r>2\rkaspi\approx\rsub$ in the \citealt{Baskin+14a} $f(r)$ profile shown in the top panel of Fig.~\ref{fig: r-distribution}, explains the apparent gap in line emission between the BLR and the NLR. 
This dusty gas at $r>\rsub$, usually referred to as the `torus', also creates the dichotomy of obscured / unobscured AGN invoked in the standard AGN unification model discussed above. 
Due to the weak line emission of the dusty gas, the physical properties of the torus have hitherto mainly been constrained via the observed IR emission and via the demographics of obscured and unobscured AGN. 
However, in Figs.~\ref{fig: r-distribution} -- \ref{fig: expected signal}
we show that this weak line emission from $r>\rsub$ can have a dramatic
effect on $S_v$ at $|v|<\hwhm$, and thus a measurement of $S_v$ can provide a new constraint on torus models.

\section{Summary}

In this paper, we have shown that the extrapolation of the $\rblr-L$ relation found by RM studies of local AGN to the most luminous quasars on the sky imply $\tblr \approx 100\muas$, with a weak dependence on $z$. 
Comparable angular sizes have previously been resolved in young stellar objects using spectroastrometry (\citealt{Pontoppidan+08,Pontoppidan+11}), suggesting that spectroastrometry is applicable also to the BLR of quasars. 

Assuming that the BLR emission has significant contributions from ordered motions, 
we calculated the expected spectroastrometric signal,  using a simple model of a rotation-dominated BLR. 
Our calculation suggests that the offset between the photocenter of the red-wing and blue-wing photons of the broad \Ha\ in luminous quasars is detectable ($\snr\approx10$) with modest time allocations on existing telescopes, and therefore can be used to constrain $\rblr$ in high-$L$ and high-$z$ quasars. 

This estimate of $\rblr$ implies a new method to spatially resolve the kinematics of the BLR and directly test whether BLR motions are ordered. It also implies a new method to contrain $\mbh$. Since previous methods to estimate $\mbh$ are based on either RM or on host properties, which work best in low-$L$ low-$z$ quasars, spectroastrometry is novel in its ability to estimate $\mbh$ in luminous quasars during the peak of the quasar epoch. 

With next-generation telescopes, BLR spectroastrometry should be routinely detectable for much fainter quasars out to $z\sim 6$, and can be expanded to include multiple broad emission lines, including weak lines, at an order of magnitude higher velocity resolution 
than achievable with existing telescopes. This will enable demographic and statistical studies of $\mbh$ and BLR properties as a function of luminosity and redshift, providing a compelling science case for next generation telescopes.

\section*{Acknowledgements}
We thank Aaron Barth for comments which significantly improved the manuscript.
We thank Eduardo Ba{\~n}ados for providing the list of high-$z$ quasars, and Alexei Baskin for providing the $f(r)$ values of \Ha\ shown in Fig.\ \ref{fig: r-distribution}.
We also thank Michael Strauss and the members of the ENIGMA group\footnote{http://www.mpia-hd.mpg.de/ENIGMA/} at the Max Planck Institute for Astronomy (MPIA) for helpful discussions. 
J.F.H.\ acknowledges generous support from the Alexander von Humboldt foundation in the context of the Sofja Kovalevskaja Award. The Humboldt foundation is funded by the German Federal Ministry for Education and Research.

\bibliographystyle{apj}

\appendix

\begin{table*}[b]
\begin{tabular}{l|c|c|c|c|c|c|c}
Object & $z$ & $L_{1450}^{(1)}$ & Line & Band & EW$^{(2)}$ & $\Phi^{(3)}$         & $\fwhm^{(4)}$  \\
       &     & $(\ergs)$        &      &      & ($\AA$)    & $(\m^{-2} \hr^{-1})$ & $(\kms)$ \\
\hline
                  3C~273 & 0.16 & 46.3 &     \Pa\ & K & 120 & $2.8 \cdot 10^{7}$  & 3400$^{(5)}$ \\
SDSS~J163302.66+234928.5 & 0.82 & 46.9 &     \Ha\ & J & 560 & $1.8 \cdot 10^{7}$  & 5050$^{(6)}$ \\
             PG~1634+706 & 1.33 & 47.5 &     \Ha\ & H & 560 & $2.7 \cdot 10^{7}$  & 4100$^{(7)}$ \\ 
SDSS~J152156.48+520238.5 & 2.21 & 47.7 &     \Ha\ & K & 560 & $1.6 \cdot 10^{7}$  & 9350$^{(6)}$ \\ 
SDSS~J155152.46+191104.0 & 2.85 & 47.8 &  \mgiip\ & J &  33 & $1.7 \cdot 10^{6}$  & 7730$^{(8)}$ \\ 
            HS~0857+4227 & 3.29 & 47.6 &  \mgiip\ & J &  33 & $7.9 \cdot 10^{5}$  & 4430$^{(9)}$ \\ 
          PSS~J1347+4956 & 4.51 & 47.6 &  \mgiip\ & H &  33 & $5.2 \cdot 10^{5}$  & 8940$^{(9)}$ \\ 
SDSS~J001115.23+144601.8 & 4.97 & 47.5 &  \mgiip\ & H &  33 & $3.5 \cdot 10^{5}$  & 5930$^{(10)}$ \\
SDSS~J114816.64+525150.3 & 6.42 & 47.1 &  \mgiip\ & K &  33 & $8.6 \cdot 10^{4}$  & 5930$^{(10)}$ \\
ULAS~J112001.48+064124.3 & 7.08 & 46.6 &  \mgiip\ & K &  33 & $2.5 \cdot 10^{4}$  & 4410$^{(11)}$ \\
\end{tabular}
\caption{Properties of the spectroastrometry candidates shown in Fig.\ \ref{fig: SN vs. z}. 
Notes: 
(1) based on $m_i$ from \S\ref{sec: angular size} and a \cite{Richards+06} SED; 
(2) EW(\Ha) from \cite{SternLaor12a}, EW(\mgiip) from \cite{VandenBerk+01}, EW(\Pa) estimated from fig.\ 1 in \cite{Petrov+12};  
(3) calculated from the line EW and a \cite{Richards+06} quasar SED normalized by $L_{1450}$;
(4) except in 3C~273, all values are estimates;
(5) from \cite{Petrov+12};
(6) based on \fwhm(\mgiip) from \cite{Shen+11};
(7) based on \fwhm(\mgiip) from \cite{MathewsWampler85}; 
(8) based on \fwhm(\civp) from \cite{Paris+14};
(9) based on \fwhm(\civp) from \cite{Shen+11}; 
(10) average $\fwhm$ of other objects;
(11) based on \fwhm(\mgiip) from \cite{DeRosa+14}
}
\label{tab: snr}
\end{table*}

\end{document}

%% file: commands.tex
\usepackage{gensymb}

\newcommand{\appropto}{\mathrel{\vcenter{
  \offinterlineskip\halign{\hfil$##$\cr
    \propto\cr\noalign{\kern2pt}\sim\cr\noalign{\kern-2pt}}}}}

\def\app#1#2{%
  \mathrel{%
    \setbox0=\hbox{$#1\sim$}%
    \setbox2=\hbox{%
      \rlap{\hbox{$#1\propto$}}%
      \lower1.1\ht0\box0%
    }%
    \raise0.25\ht2\box2%
  }%
}

\renewcommand{\th}{$^{\rm th}$}

\ifdefined\kms
\renewcommand{\kms}{\,\rm km\ s^{-1}}
\else
\newcommand{\kms}{\,\rm km\ s^{-1}}
\fi

\newcommand{\ergs}{\,\rm erg\ s^{-1}}

\ifdefined\kms
\renewcommand{\mag}{\,\rm mag}
\else
\newcommand{\mag}{\,\rm mag}
\fi

\newcommand{\mJy}{\,{\rm mJy}}

\let\AAold\AA
\renewcommand{\AA}{\text{\AAold}}
\newcommand{\mic}{\,\mbox{$\mu$m}}
\newcommand{\cm}{\,{\rm cm}}
\newcommand{\m}{\,{\rm m}}
\newcommand{\pc}{\,{\rm pc}}

\newcommand{\s}{\,{\rm s}}

\newcommand{\hr}{\,{\rm hr}}

\newcommand{\msun}{\,{\rm M_{\odot}}}

\newcommand{\muas}{\,\mu {\rm as}}
\newcommand{\mas}{\,{\rm mas}}

\def \arcsec    {^{\prime\prime}}

\renewcommand{\d}{{\rm d}}

\newcommand{\fwhm}{{\rm FWHM}}
\newcommand{\hwhm}{{\rm HWHM}}
\newcommand{\snr}{{\rm S\slash N}}

\DeclareRobustCommand{\ion}[2]{%
\relax\ifmmode
\ifx\testbx\f@series
{\mathbf{#1\,\mathsc{#2}}}\else
{\mathrm{#1\,\mathsc{#2}}}\fi
\else\textup{#1\,{\mdseries\textsc{#2}}}%
\fi}

\newcommand{\hei}{\text{He~{\sc i}}}
\newcommand{\heii}{\text{He~{\sc ii}}}

\newcommand{\nii}{\text{[N~{\sc ii}]}}

\newcommand{\sii}{\text{[S~{\sc ii}]}}
\newcommand{\siii}{\text{[S~{\sc iii}]}}

\newcommand{\oi}{\text{[O~{\sc i}]}}

\newcommand{\feii}{\text{[Fe~{\sc ii}]}}

\newcommand{\mgiip}{\text{Mg~{\sc ii}}}

\newcommand{\mgv}{\text{Mg~{\sc v}}}

\newcommand{\ciii}{\text{C~{\sc iii}]}}
\newcommand{\civp}{\text{C~{\sc iv}}}

\newcommand{\ariv}{\text{Ar~{\sc iv}}}
\newcommand{\arv}{\text{Ar~{\sc v}}}

\newcommand{\Ha}{\text{H$\alpha$}}

\newcommand{\Hb}{\text{H$\beta$}}

\newcommand{\Pa}{\text{Pa$\alpha$}}

\newcommand{\Brd}{\text{Br$\delta$}}
\newcommand{\Bre}{\text{Br$\epsilon$}}

\ifdefined\aap
\else
\newcommand{\aap}{A\&A}
\newcommand{\araa}{ARA\&A}
\newcommand{\apjl}{ApJ}
\newcommand{\apjs}{ApJS}
\newcommand{\apj}{ApJ}
\newcommand{\aj}{AJ}
\newcommand{\mnras}{MNRAS}
\newcommand{\pasp}{PASP}
\newcommand{\ssr}{Space Science Reviews}

\newcommand{\nat}{Nature}
\newcommand{\procspie}{SPIE}

\fi
\newcommand{\rmxaa}{RevMexA\&A}
\newcommand{\rvmp}{RvMP}
\newcommand{\nar}{NewAR}

\newcommand{\cloudy}{{\sc cloudy}}

\newcommand{\lrest}{\lambda_{\rm rest}}

\newcommand{\lledd}{L / L_{\rm{Edd}}}
\newcommand{\mbh}{M_{\rm BH}}

\newcommand{\rg}{r_{\rm g}}
\newcommand{\dL}{d_{\rm L}}

\newcommand{\rsub}{r_{\rm sub}}

\newcommand{\mi}{m_{\rm{i}}}
